\documentclass{aa}

\usepackage{todonotes}
\usepackage{graphicx}
\usepackage{txfonts}
\usepackage[]{hyperref}
\usepackage{import}
\usepackage[nolist,nohyperlinks,printonlyused]{acronym}
\acrodef{DM}{deformable mirror}
\acrodef{FLOWFS}{focal plane low-order wavefront sensor}
\acrodef{IWA}{inner working angle}
\acrodef{AO}{adaptive optics}
\acrodef{XAO}{extreme adaptive optics}
\acrodef{PSF}{point spread function}
\acrodef{WFS}{wavefront sensor}
\acrodef{LOWFS}{low-order wavefront sensor}
\acrodef{NCPA}{non-common path aberrations}
\acrodef{MSE}{mean squared error}
\acrodef{SNR}{signal-to-noise ratio}
\acrodef{RTC}{real-time controller}
\acrodef{DT}{digital twin}
\acrodef{LCO}{Las Campanas Observatory}
\acrodef{LLOWFS}{Lyot plane low-order wavefront sensor}
\acrodef{FPM}{focal plane mask}
\usepackage{todonotes}
\setuptodonotes{inline, noinlinepar, inlinewidth=5cm}
\usepackage{soul}
\usepackage{subfig}
\usepackage{stfloats}

\newcommand\thetextwidth{\expandafter\string\the\linewidth}

\begin{document}

\title{
	Model-based low-order wavefront sensing and control with a fully physical digital twin on MagAO-X\thanks{This paper includes data gathered with the 6.5 metre Magellan Telescopes located at Las Campanas Observatory, Chile.}
}

\titlerunning{A digital twin of FLOWFS on MagAO-X}
\authorrunning{Mars et al. }

\author{
	M.
	Mars\inst{1} \corrauth{academic@matthijsmars.com} \and S.
	Y.
	Haffert\inst{1,2} \email{haffert@strw.leidenuniv.nl} \and J.
	R.
	Males\inst{2} \email{jrmales@arizona.edu} \and L.
	M.
	Close\inst{2} \email{lclose@arizona.edu} \and K.
	Van Gorkom\inst{2} \email{kvangorkom@arizona.edu} \and O.
	Guyon\inst{2,3,4,5} \email{oliv.guyon@gmail.com} \and A.
	D.
	Hedglen\inst{6} \email{Alexander.Hedglen@ngc.com} \and P.
	T.
	Johnson\inst{2} \email{parkertjohnson1@arizona.edu} \and M.
	Y.
	Kautz\inst{3} \email{maggiekautz@arizona.edu} \and J.
	K.
	Kueny\inst{3} \email{jkueny@arizona.edu} \and R.
	Landman\inst{1} \email{rlandman@strw.leidenuniv.nl} \and J.
	Liberman\inst{3} \email{jliberman@arizona.edu} \and J.
	D.
	Long\inst{7} \email{jlong@flatironinstitute.org} \and M.
	Lucas\inst{2,4} \email{mileslucas@arizona.edu} \and J.
	Lumbres\inst{3} \email{jlumbres@arizona.edu} \and E.
	A.
	McEwen\inst{3} \email{edenmcewen@arizona.edu} \and A.
	McLeod\inst{8} \email{avalonmcleod@arizona.edu} \and T.
	Nguyen\inst{3} \email{tiffanytn@arizona.edu} \and L.
	A.
	Pearce\inst{9} \email{lapearce@umich.edu} \and L.
	Schatz\inst{10} \email{Lauren.schatz.1@spaceforce.mil} \and E.
	Tonucci\inst{1} \email{tonucci@strw.leidenuniv.nl} \and K.
	Twitchell\inst{3} \email{twitchell@arizona.edu} \and A.
	J.
	Weinberger\inst{11} \email{aweinberger@carnegiescience.edu} }

\institute{ Leiden Observatory, Leiden University, PO Box 9513, 2300 RA Leiden, The Netherlands\\ \and Steward Observatory, The University of Arizona, 933 North Cherry Avenue, Tucson, Arizona, USA \and Wyant College of Optical Sciences, The University of Arizona, 1630 E University Blvd, Tucson, Arizona, USA \and Subaru Telescope, National Observatory of Japan, National Institutes of Natural Sciences, 650 N.
	A‘ohoku Place, Hilo, Hawai‘i, USA \and Astrobiology Center, National Institutes of Natural Sciences, 2-21-1 Osawa, Mitaka, Tokyo, Japan \and Northrop Grumman Corporation, 600 South Hicks Road, Rolling Meadows, Illinois, USA \and Center for Computational Astrophysics, Flatiron Institute, 162 5th Avenue, New York, New York, USA \and Draper Laboratory, 555 Technology Square, Cambridge, Massachusetts, USA \and Department of Astronomy, University of Michigan, Ann Arbor, MI 48109, USA \and Starfire Optical Range, Kirtland Air Force Base, Albuquerque, New Mexico, USA \and Department of Terrestrial Magnetism, Carnegie Institution for Science, 5241 Broad Branch Road, NW, Washington, DC 20015-1305, USA }

\date{Received xx; accepted xx}

\abstract { Residual low-order aberrations and post-adaptive-optics tip-tilt jitter are major drivers of stellar leakage in small-inner-working-angle coronagraphs and limit high-contrast imaging performance.
}
{
	The aim of this work is to develop a fully physical and differentiable digital twin for the focal plane low-order wavefront sensor on MagAO-X.
	This digital twin can be used to provide accurate non-linear and fast linearised wavefront sensing and control to deal with low-order aberrations.
}
{
	We built a physics-based optical model of the focal plane low-order wavefront sensor system and calibrated it using laboratory and on-sky measurements.
	The calibrated model was used for non-linear phase retrieval, deriving a model-based linear reconstructor and generating the corresponding reference image for real-time control.
}
{
	The calibrated digital twin accurately reproduces the sensor response and recovers the dominant low-order aberrations.
	In the laboratory, non-linear reconstruction achieves residual floors of about 6 to 7 nm over the pupil, compared with about 11 nm for the linear reconstructor.
	The same digital-twin framework transfers across different coronagraph and wavelength configurations.
	On-sky, the model can be recalibrated with limited changes, achieves reconstruction performance close to the residual atmospheric floor, and provides a linear reconstructor that enables stable closed-loop low-order control at kilohertz speeds.
}
{
	A fully physical digital twin can unify accurate low-order wavefront reconstruction, fast linearised control, and transfer both across observing configurations and from lab to on-sky operation within a single interpretable framework.
}


\maketitle
\nolinenumbers
\section{Introduction}

High-contrast imagers rely on a combination of coronagraphy, wavefront control, and post-processing to directly image faint companions around bright objects, such as exoplanets around their host stars \citep[e.g.][]{vancapelleveenWIdeSeparationPlanets2025,closeWideSeparationPlanets2025}.
Coronagraphs block as much of the on-axis stellar light as possible to reduce the intensity difference between the star and its companion and increase contrast --- often defined as the ratio of intensity between the star and its companion \citep{kenworthyHighContrastCoronagraphy2025}.
Wavefront control through \ac{AO} is necessary to correct for aberrations in the wavefront caused by turbulence in the atmosphere or vibrations in the telescope or instrument \citep{guyonExtremeAdaptiveOptics2018}.
Keeping the wavefront as flat (minimum amount of aberrations) as possible ensures optimal coronagraph performance.
Additionally, stabilising the stellar \ac{PSF} on the detector makes it easier for post-processing techniques to remove remaining stellar light.
While \ac{XAO} systems greatly reduce the aberrations in the wavefront, their corrections are not perfect, causing stellar leakage through the coronagraph and degrading the contrast.
Additionally, vibrations introduced by the instrument behind the \ac{WFS} cannot be sensed and cause tip-tilt jitter of the stellar PSF on the coronagraph mask.
These (mostly low-order) residual aberrations are the main contribution to stellar leakage and thus loss of contrast for coronagraphs with a small \ac{IWA} that are highly sensitive to precise tip-tilt alignment of the stellar PSF on the mask \citep[e.g.][]{elsahubyPostcoronagraphicTiptiltSensing2015}.
Since the achieved contrast scales with the square of the residual wavefront error \citep[][]{malbetHighDynamicRange1995,ruaneVortexCoronagraphsHabitable2018}, correcting these residual aberrations can greatly improve contrast.

To improve the performance of small \ac{IWA} coronagraphs, a dedicated secondary \ac{WFS} can be used to measure residual atmospheric aberrations as well as aberrations caused by vibrations at high speeds.
This is often used in combination with a dedicated coronagraphic \ac{DM} that is separate from the primary \ac{WFS} and \ac{AO} control.
One option for sensing these wavefront errors is to use the rejected stellar light from a reflective \ac{FPM}, \ac{FLOWFS} \citep[][]{guyonCORONAGRAPHICLOWORDERWAVEFRONT2009}, or from a reflective Lyot stop, \ac{LLOWFS} \citep[][]{singhLyotbasedLowOrder2014,singhLyotbasedLowOrder2014a,singhOnSkyDemonstrationLowOrder2015,singhDemonstrationVersatileLoworder2017} in the case of a transmissive \ac{FPM} (e.g. a vortex coronagraph).

Current \ac{LOWFS} control algorithms typically use empirical reconstruction matrices derived from on-sky calibration.
Probing the \ac{DM} modes measures the response matrix that maps changes in \ac{DM} coefficients to changes in the wavefront-sensor signal.
A linear control matrix is then derived by inverting this response matrix.
These calibrations must be performed under observing conditions (on-sky) and calibrate the response around a static point.
They cannot be used to sense absolute wavefront aberrations but only relative changes around this static point.
Since these calibrations are performed on-sky, they are typically noisy because residual atmospheric aberrations degrade the signal, resulting in limited performance and a higher noise floor.
Because this approach assumes a linear behaviour around the static zero-point while wavefront aberrations cause a non-linear response, it is most accurate in a small region around this point.
Aberration estimation accuracy decreases for larger aberrations.
This is usually not a problem for real-time control under good seeing conditions, as the aberrations can be minimised through continuous improvement of the wavefront error.
However, if the aberrations evolve faster than the convergence of this linear control algorithm, performance is significantly reduced.

To remedy this, one can use more accurate, non-linear optimisation algorithms for wavefront sensing to better handle larger aberrations, resulting in an improved dynamic range of the controller.
However, the computational cost of non-linear optimisation is typically much higher, making it difficult to perform real-time correction at kilohertz speeds, which is necessary to control vibrations and rapidly evolving aberrations.
One option for fast non-linear solutions is a data-driven approach that uses neural networks to learn an efficient mapping from wavefront sensor images to aberrations or \ac{DM} corrections.
This approach has been applied to primary \ac{AO} \citep[e.g.][]{landmanMakingUnmodulatedPyramid2024,landmanMakingUnmodulatedPyramid2025} as well as \ac{LLOWFS} systems \citep[e.g.][]{allanDeepResidualLearning2020}.
While neural networks enable efficient and fast non-linear control, challenges remain in composing the right training data for robust performance across a range of observing conditions and instrument configurations.
Such models often require on-sky (re-)training \citep[e.g.][]{nousiainenFocalPlaneWavefront2026}.

Alternatively, one can use a model-based approach that simulates optical propagation through the instrument.
A calibrated model of a built system that accurately mirrors or forward-models the response of the instrument is often referred to as a \ac{DT}.
For these methods, the challenge lies in accurately modelling the exact response of the instrument, including all the physics that govern its behaviour, and gathering specific calibration data to determine system parameters.
The benefit of a \ac{DT} is that the model is explainable and easily adaptable as it describes the instrument across different configurations, such as coronagraphs and filters.
A fully physical \ac{DT} has the benefit of being more robust to changes in observing conditions or instrument setup as it includes strong implicit physical priors.
A fully differentiable \ac{DT} enables non-linear optimisation of its parameters and can also be useful beyond wavefront sensing and control, for example, for inverse design and forward modelling of the stellar \ac{PSF} for post-processing.

The system studied in this article is the \ac{FLOWFS} on MagAO-X \citep{malesMagAOXCommissioningResults2024}, an \ac{XAO} high-contrast imager installed on the Magellan Clay 6.5m telescope at the \ac{LCO}.

In this work, we present a fully physical and differentiable \ac{DT} for low-order wavefront sensing and control using the \ac{FLOWFS} on MagAO-X.
We show that this \ac{DT} accurately reproduces the response of the low-order wavefront sensor and can recover and correct the low-order aberrations that cause stellar leakage.
Because the model is fully physical, the framework can be adapted across different instrument configurations, including different coronagraphs and wavelength bands, with only minimal recalibration.
Because it is fully differentiable, we can perform efficient non-linear optimisation over both instrument parameters and wavefront aberrations for calibration and \ac{PSF} sharpening.
We also derive a fast linearised reconstructor for real-time low-order control at kilohertz speeds.
Finally, we demonstrate that this framework transfers from laboratory calibration to on-sky calibration and control, showing that a calibrated \ac{DT} can bridge simulation, lab, and on-sky operation.

In \autoref{sec:methods}, we describe the layout of MagAO-X, the \ac{FLOWFS} \ac{DT}, the non-linear reconstruction procedure, the calibration strategy, and the derivation of the model-based linear reconstructor used for real-time control.
In \autoref{sec:results}, we first discuss validating the \ac{DT} in the laboratory, then show its robustness across instrument configurations, demonstrate fast linearised control in the lab, and finally present its transfer to on-sky calibration, reconstruction, and control.
We conclude in \autoref{sec:conclusion} by summarising our main results and discussing the broader implications of fully physical \ac{DT}s for wavefront sensing and control.

\section{Methods}\label{sec:methods}
In this section, we describe the \ac{FLOWFS} \ac{DT} and how it is calibrated in both laboratory and on-sky scenarios.
We also describe in detail how we used non-linear optimisation techniques to recover and calibrate our model parameters.
Using this calibrated model, we derived a model-based linear interaction matrix and reference image that can be used for fast low-order control at kilohertz speeds.
While the non-linear reconstruction approach is more accurate, optimisation is slower than linear control algorithms.
Therefore, the non-linear approach was used only for model calibration and \ac{PSF} sharpening.

\begin{figure*}[!bp]
	\centering
	\includegraphics[width=\textwidth]{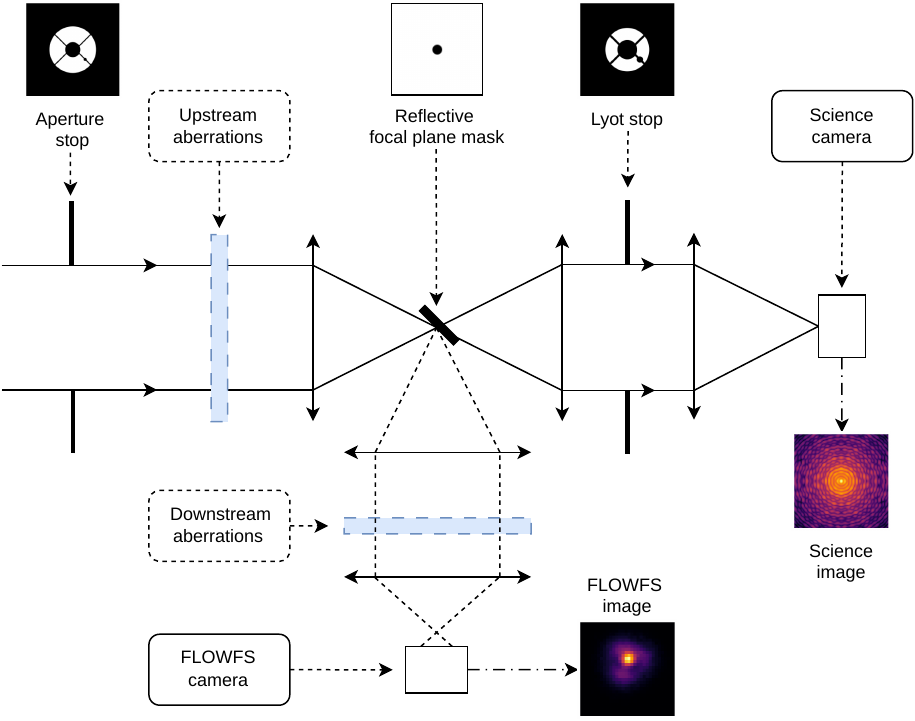}
	\caption{
		Schematic overview of the FLOWFS imaging system and \ac{DT}.
		The on-axis stellar light is blocked and reflected to a dedicated \ac{FLOWFS} camera.
		The wavefront is affected by upstream aberrations, which include both static aberrations and varying aberrations caused by residual turbulence and vibrations.
		Additionally, aberrations appear downstream of the focal plane mask.
		These aberrations are mostly static but slowly evolve over time.
		The \ac{DT} models the optical path from the aperture stop to the \ac{FLOWFS} camera, including the introduced wavefront aberrations.
		These aberrations were recovered using a phase retrieval algorithm.
		It also shows the shape of the aperture stop, the small reflective Lyot focal plane mask, and the Lyot stop.
		We also show simulated examples of a coronagraphic science image and an image from the FLOWFS camera.
	}
	\label{fig:flowfs-diagram}
\end{figure*}
\subsection{MagAO-X layout}
MagAO-X consists of an upper and lower optical bench connected by a periscope.
The instrument accepts an $f/11$ beam on the upper bench, where the wavefront is first corrected by a woofer-tweeter pair consisting of an ALPAO-97 \ac{DM} and a Boston Micromachines 2K \ac{DM} \citep{malesMagAOXProjectStatus2018,closeOpticalMechanicalDesign2018,malesMagAOXCurrentStatus2022,malesMagAOXCommissioningResults2024}.
The beam is then relayed to the lower bench, where it is split between the pyramid wavefront sensor and the science arm.
In the science arm, a Boston Micromachines 1K non-common path correction (NCPC) \ac{DM} is used for low-order \ac{NCPA} control before the beam propagates through the coronagraph optics, including the pre-apodiser, \ac{FPM}, relay optics, and Lyot stop.
The final science focus is formed at approximately $F/69$.
Depending on the coronagraph configuration, rejected stellar light from either a reflective \ac{FPM} or a reflective Lyot stop is directed to dedicated low-order sensing arms, corresponding to \ac{FLOWFS} and \ac{LLOWFS}.
These sensors are used to sense low-order aberrations that are then corrected with the NCPC \ac{DM}.

In this work, we focused on modelling and control using the \ac{FLOWFS} and considered two coronagraph configurations: a small Lyot coronagraph ($62.5$\,mas) and a large Lyot coronagraph ($104$\,mas).
We also considered observations in the z band (central wavelength $908$\,nm, $\sim14\%$ bandpass) and the r band (central wavelength $615$\,nm, $\sim18\%$ bandpass).

\subsection{Modelling \ac{FLOWFS}}
To model the \ac{FLOWFS} system, we used an optical model written in the HCIPy framework \citep{porHighContrastImaging2018}.
The code for this model can be found at \url{https://github.com/MatthijsMars/flowfs}.
\autoref{fig:flowfs-diagram} shows a schematic overview of the \ac{FLOWFS} system, which consists of several components that we need to model.
The optical model takes an input wavefront and propagates it through the system and gives the output on the wavefront sensor camera.
The model consists of several parts:
\begin{enumerate}
	\item the input aperture,
	\item upstream phase aberrations,
	\item propagation on a reflective focal plane mask,
	\item downstream phase aberrations,
	\item imaging on the detector.
\end{enumerate}
Most parameters, such as the focal lengths, physical sizes of the aperture stop and FPMs, and pixel size, were set to match the specifications of the MagAO-X design.
Other parameters, such as the focal length and magnification of the imaging on the \ac{FLOWFS} camera as well as the size of the pupil on the \ac{DM}, were determined by doing dedicated lab calibrations using the internal source of MagAO-X.
In this model, we limit ourselves to monochromatic light and phase aberrations only, which proves sufficient for this particular low-order wavefront sensor.
However, propagation over a range of wavelengths as well as amplitude aberrations can be included in such an optical model as well.

Decisions made for the particular setup of the model were based on the small Lyot coronagraph in the MagAO-X instrument.
This informed the number of sensed modes and the amount of intentional defocus for the FLOWFS camera.
Additionally, all static system parameters determined through dedicated lab experiments were fitted on data using this coronagraph.
This FPM is $272\,\mu\text{m}$ in size (corresponding to $\sim2 \lambda/D$ in the z band), and this mask acts as a spatial filter, limiting the amount of spatial information we can sense using our \ac{LOWFS}.
Therefore, in our model, we limited the upstream aberrations to nine Zernike modes (the first nine modes, excluding piston, Noll indexing).
These upstream aberrations model the static aberrations in the instrument as well as the varying residual atmospheric aberrations.
The downstream aberrations model the \ac{NCPA} between the main optical beam and the \ac{FLOWFS} beam, including aberrations due to imperfect alignments in the FLOWFS beam.
These were modelled using five Zernike modes (tip, tilt, defocus, vertical astigmatism, and oblique astigmatism).
The alignment of the FPM and the \ac{FLOWFS} camera was captured in the upstream and downstream tip and tilt phase aberrations.
Our \ac{FLOWFS} camera was also purposefully defocused by offsetting the camera by $10.5$\,mm to break degeneracies in the response to the different Zernike modes (corresponding to roughly 450\,nm root mean square of focus offset).
This defocus was also fitted in the downstream aberrations and modelled through a defocus term in the pupil.
Yet, it could also be modelled using Fresnel propagation methods for more accuracy at the cost of additional computation time.
In our model of FLOWFS, we did not require Fresnel propagation to achieve the required accuracy.
For the detector, our model sub-sampled the pixels by a factor of two to account for sub-pixel effects.
The final image was binned to $32\times32$ pixels, to match the size of the region of interest on our detector, and we also fitted one scalar parameter for the final normalisation of the forward-modelled image.
Additionally, we added a small static projection effect that models the elongation of the FLOWFS PSF along one axis ($\sim 8\%$), which is introduced in the re-imaging optics.

For each operation in the optical propagation, we can express an adjoint operation, allowing us to analytically calculate the gradient of the detector response to changes in both the upstream and downstream aberrations via back-propagation.
A detailed overview on algorithmic differentiation for phase retrieval as well as all the description of all the forward and corresponding adjoint operations used in our model can be found in \citet{jurlingApplicationsAlgorithmicDifferentiation2014}.
Alternatively, one could either calculate the gradients numerically using the finite difference method, which is significantly more computationally expensive and less accurate than analytically calculated gradients, or write the model in an auto-differentiable framework \citep[e.g.][]{baydinAutomaticDifferentiationMachine2018}.
The gradients of the detector response to the wavefront aberrations are necessary in order to utilise gradient-based non-linear optimisation techniques.

In this paper, we used analytically calculated gradients to add the functionality of gradient-based optimisation to our model in the \texttt{HCIPy} framework \citep{porHighContrastImaging2018}, which works well for models of limited complexity.
However, to optimise certain model parameters, this gradient derivation can become quite complex.
In this case, using a differentiable framework, such as $\partial\textrm{Lux}$ \citep{desdoigtsDifferentiableOpticsDLux2023}, is recommended.
Ongoing development of \texttt{HCIPy} also includes adding the option of different (differentiable) back-ends.

\subsection{Non-linear reconstruction}\label{sec:non-linear-reconstruction}
To recover the Zernike coefficients of the upstream and downstream aberrations, we minimised a pixel-wise correlation loss \citep[similarly to ][]{bradyPhaseRetrievalBasedWavefrontMetrology2018}:
\begin{equation}
	\mathcal{L}_{\text{correlation}}(\boldsymbol{y}_{\theta}, \boldsymbol{y}_{\text{obs}}) = 1 - \frac{\left(\sum_i  {y}_{\theta, i} \  y_{\text{obs}, i} \right)^2}
	{\left(\sum_i  y_{\theta, i}^2\right)\left(\sum_i  y_{\text{obs}, i}^2\right)},
\end{equation}
with $y_{\theta, i}$ and $y_{\text{obs, i}} $ the values of the $i$-th pixels of the forward-modelled and observed \ac{FLOWFS} images, respectively, and $\theta$ our model parameters.

Since this correlation loss is invariant to the normalisation of the images, we performed a final optimisation using a \ac{MSE} loss just to fit the normalisation constant:
\begin{equation}
	\mathcal{L}_{\text{MSE}}(\boldsymbol{y}_{\theta}, \boldsymbol{y}_{\text{obs}}) = \frac{\sum_i  ({y}_{\theta, i} - y_{\text{obs}, i  })^2}{N},
\end{equation}
with $N$ being the number of pixels in the images.
Note that this does not influence the recovered Zernike coefficients; thus, it is not strictly necessary for control purposes and is only used to fit the normalisation of the images for comparing the residuals.

One could use pixel-weighted versions of both these loss functions in order to weight the pixels by their \ac{SNR} or to exclude strangely behaving pixels.
In our own experiments, this did not significantly change the reconstruction quality when observing bright targets in good seeing conditions.

Since we have a fully differentiable optical model of our \ac{FLOWFS} system, it is possible to fit certain Zernike coefficients while keeping other coefficients fixed.
With this in mind, we find that the non-linear reconstructions converge more robustly by splitting up the optimisation into three stages \citep[similarly to][]{bradyPhaseRetrievalBasedWavefrontMetrology2018}:
\begin{enumerate}
	\item fit upstream and downstream tip-tilt using the correlation loss;
	\item fit all upstream and downstream Zernike coefficients using the correlation loss;
	\item fit the normalisation of the images using the MSE loss.
\end{enumerate}
This three-stage system is the same for every reconstruction, although for some optimisations we kept certain downstream aberrations fixed.

The three-stage optimisation was performed using a limited memory BFGS optimiser \citep{broydenQuasiNewtonMethodsTheir1967,byrdLimitedMemoryAlgorithm1995,zhuAlgorithm778LBFGSB1997}.
The computation time for recovering the full set of coefficients from one \ac{FLOWFS} image is $1.108\text{s} \pm 0.252\text{s}$ on a single CPU core (averaged over a full calibration set: nine modes, 11 pokes per mode, 20 frames per poke, as described in \autoref{sec:calibration}).
The optimisation is faster after calibration when only the upstream aberrations are recovered at $0.222\text{s} \pm 0.027\text{s}$.
Both computations could be sped up by implementing the model in GPU-accelerated code and running it on a GPU.
For calibration or post-processing purposes the optimisation can also be performed in parallel to speed up the calibration time.

\subsection{Calibration by sweeping through the Zernike modes}\label{sec:calibration}
To calibrate our \ac{DT} to the instrument, we determined the static aberrations in the system as well as the conversion of the Zernike coefficients placed on the \ac{DM} to the coefficients of the model.
To do this, we looped through all the modes individually and injected aberrations of different strengths (11 pokes from $\sim -1.5 \text{ rad}$ to $\sim 1.5 \text{ rad}$).
To average out additional aberrations introduced by bench seeing, vibrations, and uncorrected atmospheric aberrations (when calibrating on-sky), we acquired and reduced 20 \ac{FLOWFS} frames for every injection with an exposure time of 0.125\,ms per frame.
These different realisations were acquired by going through the full sweep of modes 20 times, ensuring sufficient time (roughly $0.1$\,s) between acquiring each realisation so that the additional aberrations are mostly uncorrelated.
We pre-processed the \ac{FLOWFS} frames similarly to the dedicated low-order \ac{RTC} on MagAO-X, so that forward-modelled images can be used as a drop-in replacement for \ac{FLOWFS} frames in the controller.
Therefore, all calibration images were pre-processed by subtracting the dark and normalising by the sum of all pixel intensities.

For the calibration, we assumed the downstream aberrations to be static and determined them by reconstructing all the aberrations (up- and downstream) using the non-linear reconstruction approach detailed in \autoref{sec:non-linear-reconstruction}.
We then fixed the downstream aberrations to the mean of the recovered coefficients over the full calibration set.
For the on-sky data, minor degeneracies between the up- and downstream aberrations and the large variance introduced by residual atmospheric aberrations result in noisy estimates of the downstream aberrations.
Therefore, we restricted ourselves to only fitting the downstream tip-tilt and setting the remaining downstream aberrations using a representative lab calibration.

With the downstream aberrations fixed, we recovered the upstream aberrations for the calibration sweep and averaged them over the 20 frames.
The implementation for the injected \ac{DM} Zernike coefficients has arbitrary units for the strength of the modes and should therefore be calibrated to match the scaling, sign, and direction of the physical modes in our model.
There should be a linear conversion factor between these coefficients and the recovered Zernike coefficients, though this factor varies slightly between modes due to asymmetries in the pupil.
Therefore, we fitted a linear relation between the input and recovered Zernike coefficients for each of the sensed modes.
The slope of this fitted line corresponds to the conversion factor between \ac{DM} and model coefficients, while the bias corresponds to the static \ac{NCPA} currently in the system.
A demonstration of this calibration is shown in \autoref{sec:validation}.
For on-sky calibrations, we only fitted the bias or static \ac{NCPA} in the system and used the conversion factors from the representative lab calibration.
These static \ac{NCPA} may differ from the lab calibration as they depend on the alignment of the \ac{FPM} and \ac{FLOWFS} camera, the differences in conditions such as temperature, as well as the \ac{DM} pattern used to sharpen the \ac{PSF}.

By repeatedly using this calibration approach and injecting opposite signs of the recovered static aberrations on our dedicated coronagraphic \ac{DM} (with some control gain applied), we can remove these static aberrations as much as possible and ensure the wavefront is as flat as possible.
This sharpens the \ac{PSF} (sharpened on the coronagraph) and minimises stellar leakage.

\subsection{Linear reconstruction}\label{sec:linear-reconstruction}
Using the calibrated \ac{DT}, we derived a model-based linear reconstructor, similarly to how one would construct an empirical control matrix.
Since this matrix must be calibrated around the operating point on-sky, an empirical control matrix would be contaminated by noise from residual atmospheric aberrations \citep[e.g.][]{singhOnSkyDemonstrationLowOrder2015}.
This approach has been trialled for MagAO-X in the past but never worked stably in a closed loop due to the uncorrected aberrations caused by bench seeing and atmospheric turbulence.
By contrast, our model-based approach infers the static upstream NCPA and the camera alignment from the on-sky calibration.
It then simulates the noise-free response of the instrument from which the control matrix is derived.

To obtain this model-based control matrix, we applied positive and negative pokes of 0.3 radians sequentially to each of the Zernike modes in our model to measure the simulated \ac{FLOWFS} response of our calibrated \ac{DT}.
This poke amplitude was chosen to lie within the linear response of the \ac{FLOWFS} found during calibration.
Note that since we have a differentiable \ac{DT}, the response could also be derived exactly in the operating point (with no probe amplitude); however, a more stable controller is achieved by averaging over a small linear range.
These pokes were applied around the calibrated static \ac{NCPA} (the fitted bias during calibration) to compute the response for each mode:
\begin{equation}
	J_m = \frac{\boldsymbol{y}_{+} - \boldsymbol{y}_{-}}{2 \Delta a},
\end{equation}
where each column of the response matrix, $J_m \in \mathbb{R}^{1024}$, is calculated from $\boldsymbol{y}_{+}$ and $\boldsymbol{y}_{-}$, which correspond to the flattened $32\times32$ pixel \ac{FLOWFS} response to a positive and negative poke of the $m$-th Zernike mode with poke amplitude $\Delta a$ (in model Zernike coefficient units), respectively.
The linear control matrix was obtained by inverting the response matrix with Tikhonov regularisation using the identity matrix ($0.01$ relative regularisation strength compared to the maximum singular value).
Finally, we applied diagonal scaling to the control matrix corresponding to the fitted slopes (mode-by-mode conversion factors from calibration; \autoref{sec:calibration}), such that the control matrix produces corrections scaled appropriately for the \ac{DM} space.
A reference image can be created by forward modelling the system at the chosen operating point through the \ac{DT}.
To retrieve the linear correction coefficient of each of the control modes, one should compute the difference between the \ac{FLOWFS} response and the reference image and multiply this with the control matrix.
The linear control matrix and the reference image are passed to the \ac{RTC} as a drop-in replacement for the empirical matrix and reference.
In the \ac{RTC} loop, a per-mode control gain can be applied to tune the strength of the corrections for each of the modes.
It is also possible to derive a linear control matrix from the calibrated \ac{DT} that removes the static \ac{NCPA} by applying the pokes around a flat wavefront (no aberrations).
However, we prefer to remove \ac{NCPA} using more accurate non-linear optimisation, then acquire a new calibration and control around this sharpened PSF.
Using the linear control matrix to calculate the correction is dominated by a single matrix-vector multiplication (with the control matrix of size $9\times1024$ and the flattened \ac{FLOWFS} response of size $1024$) and takes $12 \pm 3$\,ms on a single CPU core on a laptop.
For real-time control during lab or on-sky observations, this calculation is GPU-accelerated and can run at 8\,kHz (the maximum frame rate of our \ac{FLOWFS} camera).
These control speeds can be attained on bright, high \ac{SNR} targets.
Sampling at this rate allows us to correct for \ac{AO} residuals that evolve at the speed of our primary control loop (2\,kHz) as well as high-frequency vibrations and bench seeing introduced downstream of the primary \ac{WFS}.
For fainter targets, the control loop must be run more slowly to obtain enough signal to do accurate wavefront sensing.

\newcommand{\sweepfigscale}{0.948}
\newcommand{\injectionfigscale}{1.02}

\begin{figure*}[!bp]
	\centering
	\includegraphics[trim=14bp 16bp 1bp 9bp, clip, width=\textwidth]{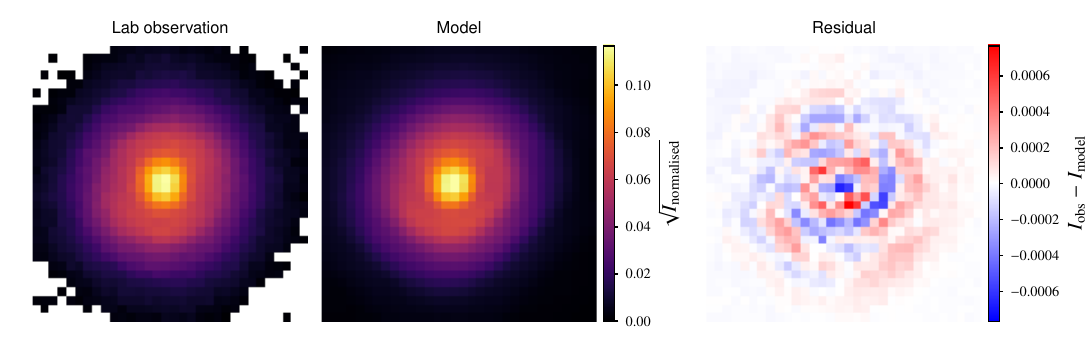}
	\caption{
		Defocused lab \ac{FLOWFS}
		PSF (left) and the reconstructed PSF (centre), using our \ac{DT} through non-linear optimisation on a square-root scale.
		The measurement was normalised by the intensity summed over all pixels.
		The residuals between the measurement and the reconstruction are shown on the right on a linear scale.
		Our model accurately captures the instrument response and recovers most wavefront aberrations introduced before and after the coronagraph.
	}
	\label{fig:psf-fit}
\end{figure*}

\section{Results}\label{sec:results}
In this section, we demonstrate that a fully physical \ac{DT} can accurately sense and control low-order wavefront aberrations, transferring from laboratory to on-sky with minimal recalibration.
First, we discuss calibrating the model and show its ability to recover wavefront aberrations in a laboratory environment.
We also show that because our \ac{DT} is fully physics-based, it is easily adapted to work in different instrument configurations.
Then, we demonstrate that this model can be used for fast control in the lab and can be calibrated for fast closed-loop control on-sky.

One thing to note is that, in all our results, we compared injected aberrations on our coronagraphic \ac{DM} with the recovered aberrations.
However, our injected aberrations are not the only aberrations in the system, as (bench) seeing introduces additional aberrations that are unknown to us.
This means that we do not have an absolute ground truth for the aberrations to be recovered.
The extent of this effect on our results is discussed in the following sections where relevant.

\subsection{Validation of the \ac{FLOWFS} digital twin}\label{sec:validation}
\autoref{fig:psf-fit} shows the non-linear reconstruction of a defocused PSF using our fully physical \ac{DT} of the \ac{FLOWFS} system, demonstrating that it accurately models the response of the instrument.
The aberrations introduced either before or after the coronagraphic mask were recovered using the non-linear optimisation procedure explained in \autoref{sec:non-linear-reconstruction}.
We recovered the downstream aberrations introduced after the coronagraph in the instrument, as well as the intentionally added defocus to the \ac{FLOWFS} camera.
The modelled \ac{FLOWFS} image closely matches the \ac{FLOWFS} response with the maximum pixel residual being $<5\%$ of the peak intensity in the observation.
The remaining discrepancies are most likely due to a combination of higher-order aberrations, simplifications in the \ac{DM} model, monochromatic modelling, and unmodelled reflections of the glass substrate that the mask sits on.
Nevertheless, the results presented in this work demonstrate that the model can accurately recover the induced aberrations and close the loop stably despite these residuals.

\begin{figure*}[!tp]
	\centering
	\includegraphics[trim=10bp 15bp 5bp 16bp, clip, scale=\sweepfigscale]{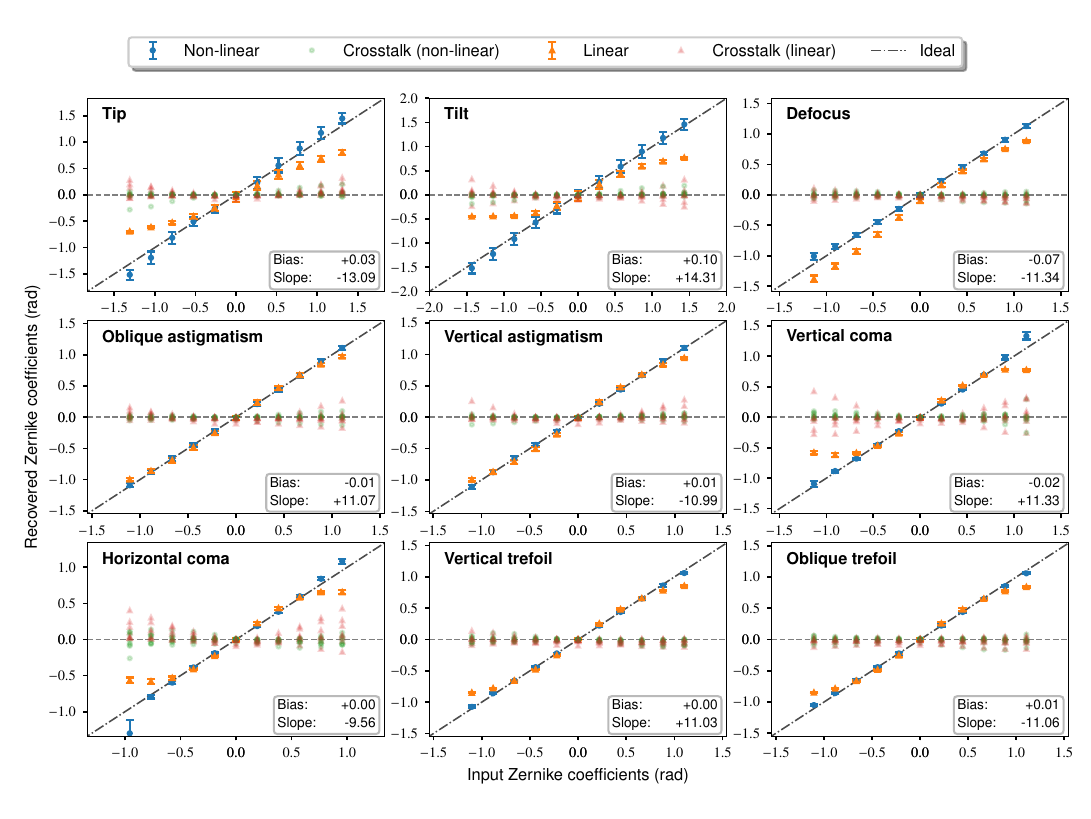}
	\caption{
		Average reconstructed Zernike coefficients, obtained using both the non-linear and linear reconstruction methods for injected aberrations of one mode at a time in the lab with the small Lyot coronagraph in the z band.
		The reconstructed coefficients were corrected by fitting the slope or conversion factor between the \ac{DM} and model Zernike coefficients and the bias that represents the static aberrations in the system, as described in \autoref{sec:calibration}.
		The error bars show the standard deviation of the reconstructed modes over 20 frames of lab data.
		This variation is mostly visible in the tip-tilt modes and is introduced by bench seeing and vibrations.
		Additionally, in each sub-plot, the crosstalk is indicated by plotting the reconstructed mode coefficients of the non-injected modes.
		The non-linear reconstructions are more accurate and show a linear response to increases in input aberration with only minor crosstalk between the tip-tilt and the corresponding coma modes.
		The linear reconstructor loses some linearity at the largest input aberrations and shows increased crosstalk between the tip-tilt and coma modes.
	}
	\label{fig:sweep-lab-lyotsm}
\end{figure*}

Second, we observe the linearity of the modal reconstruction by sweeping through the nine selected Zernike modes and reconstructing the response (see \autoref{sec:calibration}) used to calibrate the model.
\autoref{fig:sweep-lab-lyotsm} shows this modal response for both the non-linear reconstruction method and the linear reconstructor derived after model calibration, as described in \autoref{sec:linear-reconstruction}.
For the non-linear reconstructions, we get a modal response that is linear for all modes.
The variation in the reconstructed tip and tilt coefficients is slightly larger than that in the other modes due to additional aberrations introduced by vibrations.
Additionally, the sensed Zernike modes are not fully orthogonal due to the large central obscuration of the Magellan telescope ($\sim30\%$), asymmetries in the pupil from the spiders, and the illumination of the coronagraphic \ac{DM} being smaller ($\sim12\%$) than the full extent of the \ac{DM} (and the patterns applied to it).
While these effects are modelled by the \ac{DT}, this causes some crosstalk between modes, which is particularly noticeable between the coma and tip-tilt modes.
For the linear reconstructor, this crosstalk between the tip-tilt and coma modes is more pronounced.
There is also some general loss of linearity at large input aberrations, yet the response is mostly linear for most of the aberration range.
The crosstalk could be partially remedied by orthogonalising the control mode basis.
In this work, we decided to continue using the circular Zernike basis for easier interpretability and implementation.
The conversion factor from \ac{DM} to model mode coefficients (the fitted slopes of the lines used to correctly scale the input aberration coefficients) varies slightly across the coefficients.
This is due to asymmetries in the telescope pupil and the \ac{DM} not being fully illuminated.
We do see that these conversion factors also describe the response of the linear reconstructor accurately.
There are also some static aberrations in the instrument, as indicated by the fitted bias for each of the coefficients.

\begin{figure*}[!t]
	\centering
	\subfloat[]{\label{fig:injection-lab-lyot-sm-z}\includegraphics[trim=9bp 0 11.5cm 4bp, clip, scale=\injectionfigscale]{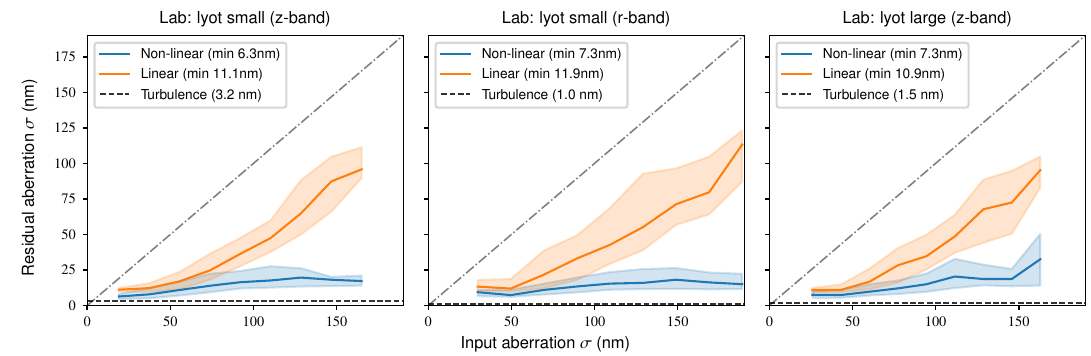}}
	\subfloat[]{\label{fig:injection-lab-lyot-sm-r}\includegraphics[trim=7cm 0 6cm 4bp, clip, scale=\injectionfigscale]{Figures/lyot_injection_lab_row.pdf}}
	\subfloat[]{\label{fig:injection-lab-lyot-lg-z}\includegraphics[trim=12.5cm 0 1bp 4bp, clip, scale=\injectionfigscale]{Figures/lyot_injection_lab_row.pdf}}

	\caption{
		Residuals between the reconstructed and input wavefront aberration, calculated as the standard deviation of the differential phase over the pupil in a laboratory environment, shown for three different instrument setups: (a) the small Lyot coronagraph ($272\,\mu\text{m}$) in the z band, (b) the small Lyot coronagraph in the r band, and (c) the large Lyot coronagraph ($453\,\mu\text{m}$) in the z band.
		All reconstructions were averaged over 20 frames to reduce the effect of bench seeing and vibrations in the system.
		The plot shows the mean (solid lines) and standard deviation (shaded areas) of the residuals for 1000 (averaged) reconstructions in ten bins containing roughly 100 samples each, as a function of the input aberration.
		The linear reconstructor reduces the wavefront aberrations but performs significantly worse for larger input aberrations.
		The non-linear reconstruction method shows only a slight increase in the residuals at larger wavelengths.
		The estimated remaining turbulence in the reconstructions after frame averaging is shown in the plots as 'turbulence'.
		The performance of the phase retrieval is very similar across the three instrument configurations.
	}
	\label{fig:injection-lab}
\end{figure*}

Figure~\ref{fig:injection-lab-lyot-sm-z} shows the model performance for reconstructing 1000 randomly injected aberrations of different amplitudes (of the nine input modes) using the small Lyot coronagraph in the z band, each averaged over 20 frames (with 0.125\,ms exposure time, separated by $\sim 1$\,s) to counteract bench seeing and vibrations.
The non-linear reconstruction method is generally more accurate than the linear reconstructor, and this difference is particularly noticeable for larger aberrations.
For control purposes, both methods reduce the aberrations significantly.
However, we note a trade-off between reconstruction speed (using a linear controller) and accuracy and thus dynamic range (using a non-linear controller).
These methods slowly reduce aberrations when run in closed loop and converge to the floor of their reconstruction performance when correcting faster than the aberrations evolve.
Since both methods reduce aberrations significantly for any input aberration, the performance floor for control using these techniques can be seen in the error achieved at the lowest input aberrations.
When wavefront control is faster than aberration evolution, the residual floor is $\sim6$\,nm over the pupil for the non-linear method and $\sim11$\,nm for the linear reconstructor.
Due to the increased computation time of the current implementation of the non-linear reconstructor, it cannot be run at kilohertz speeds.
However, it could be used when observing fainter targets, where the longer integration time required to collect the signal necessary for accurate phase retrieval is the limiting factor that sets the control rate.
In future work, we aim to significantly decrease the computation time required for the non-linear reconstructions to run them at kilohertz speeds and thus achieve a lower residual floor for control on a wider range of targets.
The residual errors quoted above include contributions from unknown bench seeing and vibrations, which make the reconstruction performance appear worse.
To estimate this effect, we also reconstructed several batches of frames with no injected aberrations.
These reconstructions were also averaged over 20 frames, and the variation in turbulence after averaging is around $\sim3$\,nm across the pupil.
This estimate also includes model uncertainty, because it is based on recovered wavefront aberrations rather than known induced aberrations, but it indicates the contribution of these unknown aberrations to the measured residuals.

\subsection{Model robustness to instrument setup configurations}\label{sec:robustness}
Since our model is fully physical, it demonstrates robust performance across changing instrument setups and can be easily adjusted by replacing their respective parts of the \ac{DT}.
Figures~\ref{fig:injection-lab-lyot-sm-r}~and~\ref{fig:injection-lab-lyot-lg-z} show the reconstruction performance when switching from the z band to the r band or from the small Lyot \ac{FPM} to the large \ac{FPM} ($453\,\mu\text{m}$).
In both cases, we adjusted the \ac{DT} by swapping in the correct coronagraph model or by changing the wavelength of the input wavefront and acquiring a new calibration measurement set.
While the current amount of introduced defocus is optimised for the small Lyot coronagraph in the z band, our model demonstrates similar performance in these other instrument configurations with the non-linear and linear reconstructions converging to $\sim7$\,nm and $11-12$\,nm, respectively, with the turbulence estimated to be $\sim1$\,nm.
There is no need to re-fit any of the system parameters, such as the focal length, magnification, DM illumination, and projection of the \ac{FLOWFS} arm.
Rather, only the static \ac{NCPA} and \ac{DM} conversion factors need to be derived, since they depend on the particular mask or wavelength used in the observation.
The fact that the only difference between instrumental setups can be explained using a difference in static wavefront aberrations demonstrates that our physics-based approach generalises across configurations without tuning and supports its physical accuracy and flexibility.

\subsection{Linearised control performance in the lab}\label{sec:lab-control}
\begin{figure*}
	\centering
	\includegraphics[trim=16bp 2bp 9bp 3bp, clip, scale=0.948]{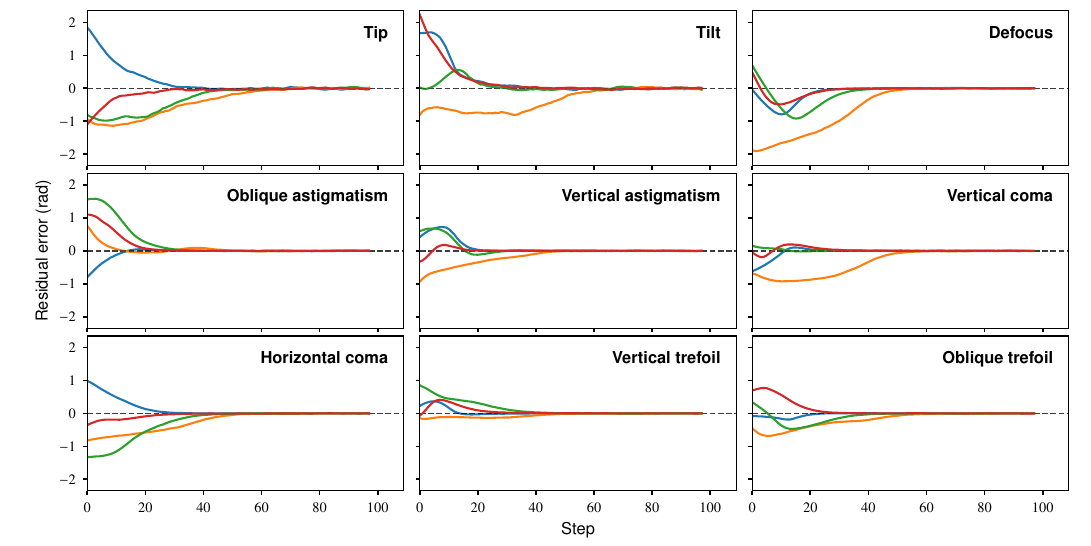}
	\caption{
		Linear control of the nine Zernike modes for the four injected aberrations indicated by the differently coloured lines.
		After 60 steps of the linear control algorithm (corresponding to $<10\text{ms}$ at 8\,kHz), all modes converge.
	}
	\label{fig:control-lab}
\end{figure*}

To demonstrate the performance of the linearised control matrix, \autoref{fig:control-lab} shows how injected aberrations ($\sim 300$\,nm wavefront aberration) are corrected by examining each of the modes.
The corrections were applied at the imaging speed of the science camera (4\,Hz), allowing us to see the changes to the PSF.
In real scenarios, however, control occurs at a much faster speed.
We see that all injected aberrations are removed in at most 60 iterations of the control loop.
This corresponds to less than 10\,ms when running the control loop at our typical operating speed (8\,kHz) for bright targets.
We observe that initial confusion between modes can occur, as indicated by some modal coefficients being adjusted further from zero aberration.
However, this crosstalk or confusion disappears as the overall wavefront error decreases, until all modes converge to their zero-point.
This behaviour is analogous to what we see in \autoref{fig:injection-lab}, where, for higher aberrations, the reconstruction quality of the linear reconstructor is poor but improves when the aberrations get smaller.
This behaviour is also somewhat expected, as the linear reconstructor is derived around the zero-point of the system and the system only behaves linearly in a limited range.
We also see that the tip-tilt corrections are less smooth due to the vibrations in the instrument, which is consistent with the behaviour seen during calibration (\autoref{sec:validation}) and does not affect the ability to converge.

\subsection{On-sky calibration and reconstruction}\label{sec:results-on-sky}
\begin{figure*}[!tp]
	\centering
	\includegraphics[trim=10bp 15bp 5bp 16bp, clip, scale=\sweepfigscale]{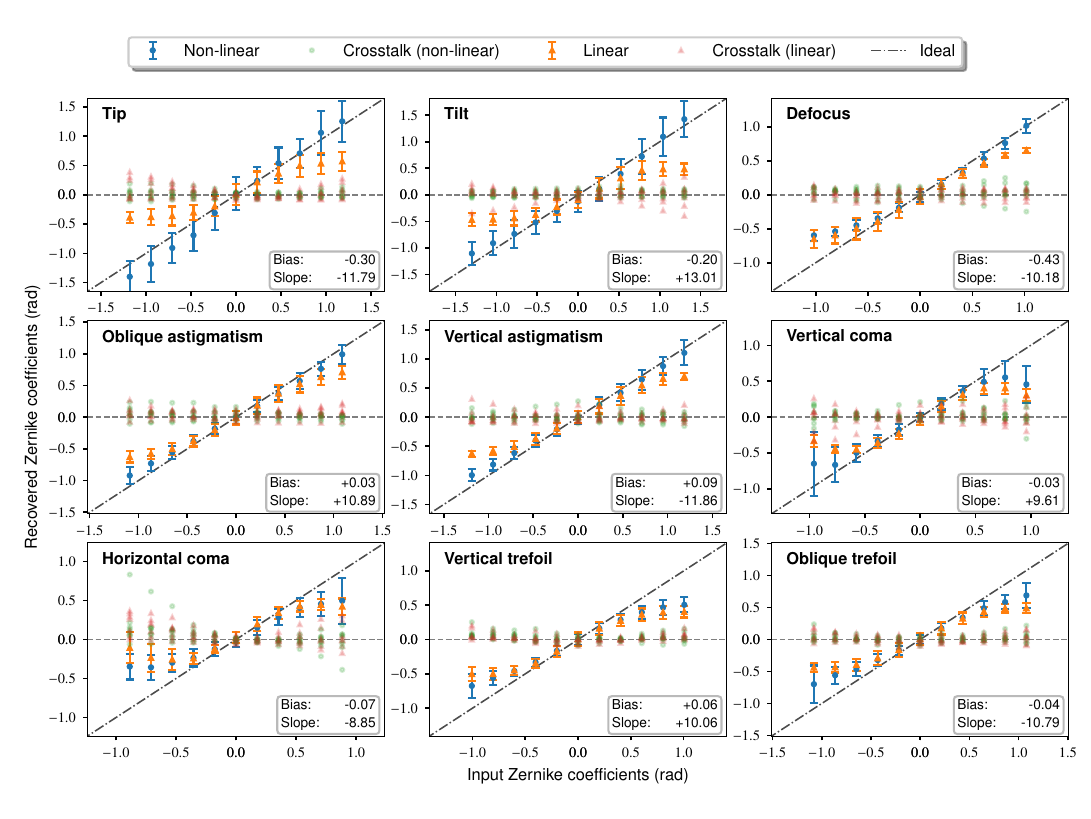}
	\caption{
		Reconstructed Zernike coefficients for the on-sky calibration with the large Lyot coronagraph in the z band for both the non-linear and linear reconstruction methods.
		All reconstructions were averaged over 20 frames to mitigate uncorrected atmospheric aberrations.
		Additionally, in each sub-plot, the crosstalk is indicated by plotting the reconstructed mode coefficients of the non-injected modes.
		The slopes converting the \ac{DM} coefficients into model coefficients were taken from a representative lab calibration and were not fitted during this on-sky calibration.
		We fitted the static aberration in the system (the bias of the lines) and used this as the zero-point for the linear controller.
	}
	\label{fig:sweep-on-sky-lyotlg}
\end{figure*}

Finally, we demonstrate that these physical \ac{DT}s are easily adapted to work on-sky.
While we wanted to test several configurations of the instrument on-sky, due to observing conditions and limited available observing time, we were only able to get a high-quality calibration dataset for the large Lyot coronagraph.
\autoref{fig:sweep-on-sky-lyotlg} shows the calibration of our \ac{FLOWFS} \ac{DT} with the large Lyot coronagraph in the z band while observing on-sky (2 December 2025, $\sim 0.5$" seeing).
During on-sky calibration, we used a lab calibration (taken two days prior) for the downstream aberrations, and we only fitted the upstream aberrations as well as the downstream tip-tilt for detector alignment (as discussed in \autoref{sec:calibration}).
The reconstructions exhibit much larger variance than in the laboratory calibrations due to the presence of unknown residual atmospheric aberrations.
In the small aberration regime, both the linear and non-linear reconstructors exhibit a good linear response.
Examining larger aberrations, we observe a deviation from linearity and an increase in crosstalk between the recovered modes, particularly between the coma and tip-tilt modes.
There is also an asymmetric response when examining the defocus term, where for negative injected aberrations it seems to flatten out.
This is due to the intentional downstream defocus being optimised for the smaller Lyot coronagraph.
When we introduce a large opposite defocus on the \ac{DM}, we enter a regime where the sensor is less sensitive to the aberrations.

\clearpage
\begin{figure}[!t]
	\centering
	\includegraphics[trim=15bp 21bp 3bp 9bp, clip, scale=\injectionfigscale]{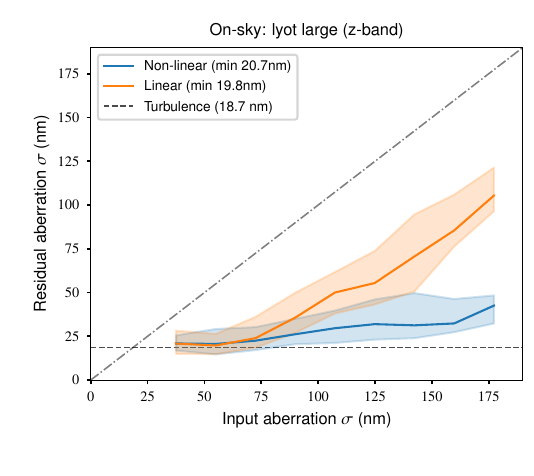}
	\caption{
		Mean (solid lines) and standard deviation (shaded area) of the residual aberration error from reconstructing input aberrations using the non-linear and linear reconstruction methods for 1000 randomly injected aberrations, plotted in ten bins containing roughly 100 samples each.
		Each individual reconstruction was averaged over 20 frames to mitigate the additional aberrations induced by atmospheric turbulence.
		The residual unknown turbulence was estimated from \ac{FLOWFS} frames with no injected aberrations.
		For each estimate, we reconstructed 20 frames individually and then averaged the reconstructions.
	}
	\label{fig:injection-on-sky-lyotlg}
\end{figure}
\begin{figure*}[!b]
	\centering
	\includegraphics[trim=8bp 1bp 17bp 5bp, clip, scale=0.92]{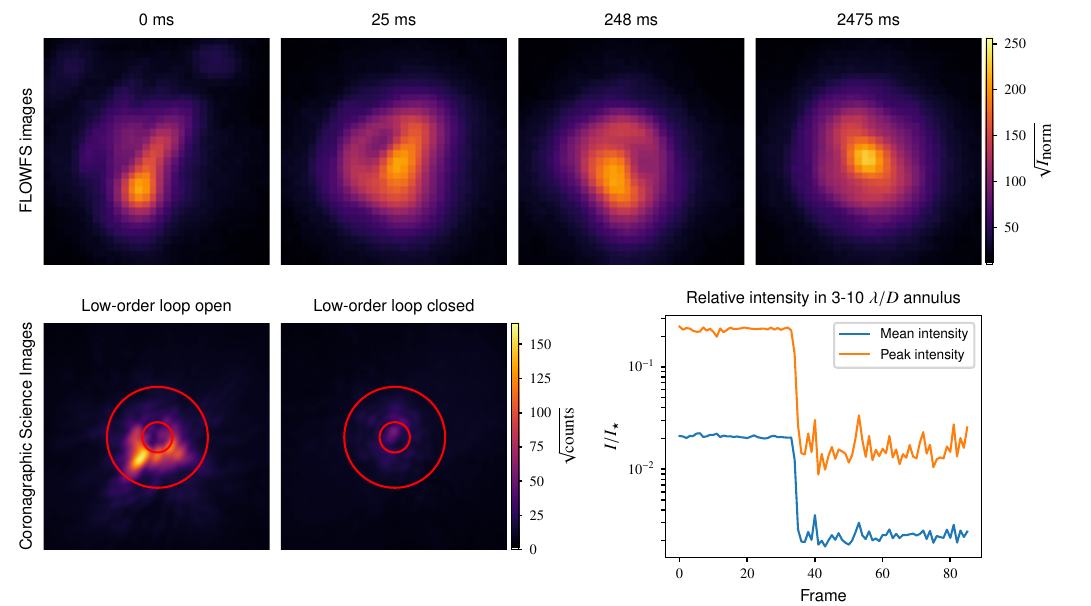}
	\caption{On-sky low-order control and its effect on stellar leakage.
		Top row: Images from the \ac{FLOWFS} camera after closing the loop.
		The \ac{DT}-based linear controller removes the aberrations and stabilises the \ac{PSF} in the presence of residual atmospheric aberrations.
		Bottom row, left: Comparison of the coronagraph images on the science camera before and after loop closure, with a $3 - 10 \lambda / D$ region indicated by two red circles.
		Bottom row, right: Mean and peak intensity in the $3-10 \lambda / D$ region over the period when the low-order control is turned on.
		Closing the loop removes the aberrations and therefore reduces leakage.
		We note increased variability after closing the loop due to sub-optimal modal control gains, resulting in a slightly overdriven system and additional tip-tilt jitter.
	}
	\label{fig:control-on-sky}
\end{figure*}

Using the calibrated \ac{DT}, we demonstrate the reconstruction performance of the non-linear and linear reconstructors on injected aberrations in \autoref{fig:injection-on-sky-lyotlg}.
Both the linear and non-linear reconstructors can reduce the input aberrations significantly and are therefore well suited to on-sky control.
Furthermore, the non-linear approach exhibits only a slight increase in residual aberration when recovering higher input aberrations, indicating that it has good and stable recovery across a large range of input aberrations.
The unknown residual atmospheric aberrations set a performance floor of $\sim18.7$\,nm (estimated from 20-frame averaged zero-injection reconstructions, with the 20 frames spread over $\sim20$\,s).
The performance floors of the non-linear and linear reconstructors are $20.7$\,nm and $19.8$\,nm, respectively.
The linear reconstructor performs slightly better in the mean residual, though the two are similar within the observed variation of the reconstructions.
In the figure, we can see that the performance floors of both reconstructors are within one standard deviation of the limit set by the unknown residual atmospheric aberrations.
However, to get a better estimate of the true performance, a significantly larger number of frames should be taken for each input aberration to average out the residual atmospheric turbulence.

\subsection{On-sky control}\label{sec:on-sky-control}
We also tested on-sky real-time control of low-order aberrations using the linear reconstructor derived from our on-sky calibrated twin.
\autoref{fig:control-on-sky} demonstrates the on-sky control of nine Zernike modes using the small Lyot coronagraph on Alpha Centauri (17 April 2025, $\sim 0.5"$ seeing).
We injected aberrations into the system by applying them to the dedicated coronagraphic \ac{DM} and then closed the low-order loop.
During this observing run, our calibration pipeline was sub-optimal, as it did not fully average out the residual atmospheric aberrations.
Despite this, the controller removed the aberrations, reducing leakage by about one order of magnitude (compared to the case with the injected aberrations), and ran stably in closed loop at 8\,kHz.
In this particular demonstration, the control gain was not set optimally.
This caused overdriving of the tip-tilt modes, which introduced additional tip-tilt jitter.
Additionally, the modal control gain on the other modes was most likely too low, which, in combination with constantly varying residual aberrations, made convergence and removal of all aberrations slow.
This is easily remedied by properly optimising the modal gains.
Most of the leakage is removed in only a few science frames, as shown in the graph of relative intensity in the $3 - 10\,\lambda/D$ region.

\section{Conclusions}\label{sec:conclusion}
Our \ac{DT} of the \ac{FLOWFS} system in MagAO-X accurately models the response of the optical system and can therefore be used for wavefront reconstruction and subsequent control.
During lab calibration, we found a highly linear relation between the recovered and injected aberrations when using the non-linear reconstructor.
There is a slight amount of crosstalk between the tip-tilt and coma modes, most likely introduced by the illumination of the beam not matching the size of the modes placed on the \ac{DM} (with the \ac{DM} patterns being larger than the beam size).
After lab calibration, the residual aberration floors are approximately 6-7\,nm for non-linear reconstruction and 11\,nm for linear reconstruction, consistent across different instrument setups.
During on-sky calibration, we estimated the residuals to be 20.7\,nm and 19.8\,nm for the non-linear and linear reconstructors, respectively, compared to the floor set by unknown residual atmospheric aberrations estimated at 18.7\,nm.
Since our \ac{DT} is fully physical, it is easily adapted to different instrument setups such as using different coronagraphs and different wavelength filters.
Using a linear controller derived from our calibrated DT, we can perform fast linear control at kilohertz speeds to stabilise the \ac{PSF} in real time.
Slower non-linear control is more accurate for larger aberrations and has an improved reconstruction floor, making it ideal for \ac{PSF} sharpening and model calibration.
We successfully calibrated our \ac{DT} on-sky and ran closed loop at 8\,kHz using the fast linear controller derived from our DT, reducing leakage and keeping the \ac{PSF} stable.

Since our \ac{DT} captures the \ac{NCPA} in the \ac{FLOWFS} branch as well as any static \ac{NCPA} in the system both upstream and downstream of the focal plane mask, the more accurate non-linear reconstruction could be used to track the evolution of the \ac{NCPA} while observing in closed loop.
If the \ac{NCPA} changes significantly, an updated model could be used to regenerate the linear control matrix and the reference image on the fly, without interrupting the science observation.

The benefits of fully physical and differentiable DTs include easy adaptation to different instrument setups and to evolving \ac{NCPA}, compared to fully data-driven approaches that require a dedicated model for each different setup.
Additionally, our model is fully explainable and can be used to diagnose misalignments in the system such as sensing the downstream \ac{NCPA}.
The fully differentiable nature of our model enables inverse design, for instance, optimising the reflective surface on the focal plane mask or the focus diversity used, in order to maximise the sensitivity of the wavefront sensing setup.

Digital twins are a promising framework for providing telemetry-based \ac{PSF} modelling \citep[e.g.][]{guyonHighContrastImaging2022}.
By building not only a model of the wavefront sensor but also a model of the science beam, we can reconstruct the wavefront aberrations in our system and forward-model the recovered aberrations through the science beam model to estimate the stellar \ac{PSF} throughout the observation, which can be used for post-processing.

In future work, we will work on providing telemetry-based \ac{PSF} modelling for MagAO-X and developing a \ac{DT} for wavefront sensing and control in \ac{LLOWFS} configurations of the instrument.
Additionally, we are currently working to reduce non-linear reconstruction time so that it can be used for fast real-time control.

\begin{acknowledgements}
	This paper includes data gathered with the 6.5 metre Magellan Telescopes located at Las Campanas Observatory, Chile.
	The MagAO-X Phase II upgrade program is made possible by the generous support of the Heising-Simons Foundation.
	We are very grateful for support from the NSF MRI Award \#1625441 (MagAO-X).
	MagAO-X uses the CACAO software package, which is supported by NSF Award \#2410616.
	MM and SYH acknowledge support from NWO Award 184.036.004.
	JL and SYH acknowledge support from NASA APRA award 80NSSC24K0288.
\end{acknowledgements}

\bibliographystyle{aa}
\bibliography{Bibliography/lib.bib}

\end{document}